\documentclass[runningheads]{svmult}
\usepackage{makeidx}
\usepackage{graphicx}
\usepackage{subeqnar}
\usepackage{multicol}
\usepackage{physprbb}
\makeindex

\begin{document}

\title*{On Multiwavelength Emission and Morphology of Large-Scale Quasar Jets}
\toctitle{On Multiwavelength Emission and Morphology of Large-Scale Quasar Jets}
\titlerunning{Emission and Morphology of Quasar Jets}
\author{\L . Stawarz\inst{1}
\and M. Sikora\inst{2}
\and M. Ostrowski\inst{1}
\and M.C. Begelman\inst{3}}
\authorrunning{\L . Stawarz et al.}
\institute{Obserwatorium Astronomiczne, Uniwersytet Jagiello\'nski, ul. Orla 171, 30-244 Krak\'{o}w, Poland
\and Centrum Astronomiczne im. M. Kopernika, ul. Bartycka 18, 00-716 Warszawa, Poland
\and Joint Institute for Laboratory Astrophysics, University of Colorado, Boulder, CO 80309-0440, USA}

\maketitle

\begin{abstract}
We propose, that the knots of the large-scale jets in powerful radio sources represent moving and separate portions of the jet matter, with the excess kinetic power. This can explain many morphological properties of quasar jets --- like high knot-to-interknot brightness contrasts, frequency-independent knot profiles and almost universal extents of the knot regions --- independently of the exact emission mechanisms responsible for producing broad-band jet emission. We briefly discuss a possible connection between this scenario and the idea of highly modulated/intermittent activity of the jet engine.
\end{abstract}

Growing numbers of large-scale jets in powerful extragalactic radio sources are now being observed at optical and X-ray frequencies (for a recent review see~\cite{sta04a}). The high angular resolution of {\it Hubble Space Telescope} and {\it Chandra X-ray Observatory} coupled with {\it VLA} results have given us the unique opportunity to analyze also the morphological characteristics of jets at different photon energies. The implied properties, like (i) knotty morphology with a high knot-to-interknot brightness contrast, (ii) frequency-independent knots' profiles and similar extents of the knots when observed at different photon energies ($\sim 1$ kpc), (iii) lack of correlation between spectral and brightness changes along the jet, and (iv) spatial offsets between the maxima of the X-ray and radio emission within the knot regions, have to be explained by any model regarding multiwavelength emission of the large-scale quasar jets.

All these characteristics are difficult to understand if the knots are identified with strong stationary (e.g., reconfinement) shocks through which the jet matter flows continuously. As discussed in~\cite{sta04}, other possibilities (like the `clumping model',~\cite{tav03}) are also in many aspects problematic. In Stawarz et al. (\cite{sta04}) we proposed instead, that the knots represent portions of a relativisticaly moving jet with excess kinetic power. Such a situation is expected if the activity of central engines in radio-loud AGNs is intermittent, or highly modulated. Then, many morphological properties of the large-scale quasar jets can be explained in a natural way, almost independently of the exact emission processes involved in producing the jet broad-band spectra.

The idea of intermittent jet activity was considered previously in the context of `partial' radio jets associated with the radio galaxies 3C 219 and 3C 288~\cite{bri89,cla92}, in connection with extended radio emission observed in a few GPS sources~\cite{bau90,sta90}, and regarding some morphological features observed in Giant Radio Galaxies~\cite{sub96} including Double-Double Radio Galaxies~\cite{sch00}. Also, the observed distribution of sizes for extragalactic radio sources, showing an overabundance of compact radio objects when compared with the number of classical doubles, led Reynolds \& Begelman (\cite{rey97}) to consider intermittent character of the jet phenomenon.

The appropriate time scales for jet high activity and quiescent epochs required to explain morphological properties of the large-scale quasar jets considered here are, respectively, $\sim 10^4$ yr and $\sim 10^5$ yrs, as the linear sizes of the knot regions are $\sim 1$ kpc while typical distances between the knots are on average an order of magnitude longer (similar time-scales were considered in~\cite{rey97}). This, together with the month-to-year variability of the jets at small (parsec) scales, and evidences for the restarting jet activity on the time scale of $10^{7} - 10^{8}$ yrs in the Double-Double Radio Galaxies, indicate that the jet activity in powerful radio sources is variable/modulated/intermittent over many different time-scales (see in this context~\cite{sta04b} for the special case of quasar 3C 273).

It can be noted that $10^4$-year-long active epoch separated by the longer periods of quiescence can appears also in the thermal-viscous instability model for the AGN accretion disks~\cite{sie97}. Thus, the jet activity, and hence the whole evolution of the radio source, could be driven by unstable accretion disks surrounding supermassive black holes. We also note that Galactic X-ray Transients are also known to produce radio emission which is widely modeled in terms of multiple jet-like ejection events. Modulated jet activity of the large-scale jets considered here constitutes an interesting link between quasar and microquasar phenomena, although the physical processes controlling intermittent/highly modulated activity of the central engines in both types of sources can be different. As a result, the time scales for the active and quiescent periods in galactic and extragalactic jet sources do not necessarily scale in a simple way with the mass of the central compact object.\\
\\
\L S, MS and MO were supported by the grant PBZ-KBN-054/P03/2001. MCB acknowledges support from National Science Foundation grant AST-0307502.

\end{document}